\documentclass[english,aps,prl,amsmath,amssymb,twocolumn,letterpaper,citeautoscript,superscriptaddress,longbibliography]{revtex4-1}

\usepackage{mathtools}

\usepackage[normalem]{ulem} 

\usepackage{comment}
\usepackage{amsmath}
\usepackage{physics}
\usepackage{graphicx}
\usepackage{babel}
\usepackage{amstext}
\usepackage{esint}
\usepackage[bookmarks=false,linkcolor=blue,urlcolor=blue,colorlinks,citecolor=blue]{hyperref}
\usepackage{pdfpages}	
\usepackage{pgffor}

\makeatletter
\AtBeginDocument{\let\LS@rot\@undefined} 
\makeatother

\begin{document}

\title{Statistical Signatures of Majorana Zero Modes in Disordered Topological Superconductor Antidot Vortices}

\author{Zhibo Ren}

\affiliation{Department of Physics and Astronomy, Purdue University, West Lafayette, Indiana 47907 USA}

\author{Jukka I. V\"ayrynen}

\affiliation{Department of Physics and Astronomy, Purdue University, West Lafayette, Indiana 47907 USA}

\date{\today}
\begin{abstract}
An antidot-pinned vortex in a three-dimensional topological insulator-superconductor platform hosts a Majorana zero mode (MZM). However, numerous Caroli-de Gennes-Matricon (CdGM) states coexist with it. We develop a general theory to study the effects of disorder on the system, emphasizing the difference between Majorana zero mode and CdGM states. Using both an analytical random matrix theory approach and numerical simulations, we derive the statistical distributions of these states. Our results demonstrate that the variance of the MZM probability density is twice that of the CdGM states, a difference due to the former having a real wave function as opposed to a complex one.  
This distinction can be measured by using scanning tunneling microscopy in a disordered antidot vortex, providing a signature of MZM beyond the zero-bias conductance peak. 
\end{abstract}
\maketitle

\textbf{\emph{Introduction.~}}
Majorana zero modes (MZMs) in topological superconductors have attracted sustained interest due to their non-Abelian exchange statistics and potential applications in fault-tolerant quantum computation~\cite{PhysRevLett.86.268,2003AnPhy.303....2K,RevModPhys.80.1083}. While MZMs have been predicted in various platforms such as topological superconductor nanowires~\cite{PhysRevLett.105.077001,PhysRevLett.105.177002,PhysRevLett.104.040502,PhysRevB.82.094522,DasSarma2023}, topological insulator-superconductor heterostructures~\cite{PhysRevLett.100.096407},  fractional quantum Hall edges~\cite{MOORE1991362,PhysRevB.61.10267}, and vortices in Fe-based superconductors~\cite{Qin_2019,10.1093/ptep/ptad084,PhysRevB.103.L140502,PhysRevLett.129.277001}, their unambiguous identification remains challenging. In two-dimensional (2D) spinless p-wave superconductors, vortices provide a natural platform for hosting MZMs~\cite{PhysRevB.61.10267,PhysRevLett.86.268,volovik1999,PhysRevB.70.205338,PhysRevB.73.014505,PhysRevLett.100.096407}. However, the same vortex also supports   Caroli–de Gennes–Matricon (CdGM) states that arise generically in vortex cores~\cite{CAROLI1964307}. These low-energy states are closely spaced in energy and spatially localized near the vortex, with disorder potentially driving them to lower energies~\cite{PhysRevB.86.035441,PhysRevB.110.075433,PhysRevLett.130.106001,PhysRevB.107.184509}, complicating both theoretical characterization and experimental detection of the MZM, similar to the situation in nanowires~\cite{PhysRevLett.109.227005,PhysRevLett.109.267002,PhysRevB.97.165302,Prada2020,Yu2021,Valentini2021,PhysRevB.107.184509,PhysRevB.106.035413}. 

In realistic systems, disorder is unavoidable and modifies the spatial structure and energy distribution of both MZMs and CdGM states~\cite{PhysRevLett.107.196804,PhysRevLett.109.227005,RevModPhys.87.1037,PhysRevB.107.184509,PhysRevB.109.L180506,doi:10.34133/research.1087}. Understanding how the two types of states behave under disorder is therefore essential for interpreting tunneling spectroscopy measurements, particularly scanning tunneling microscopy (STM) experiments that probe the local density of states inside vortex cores~\cite{PhysRevLett.62.214,PhysRevB.106.035413,chen2018discrete,doi:10.1126/science.aao1797,PhysRevX.8.041056,Chen_2019,Machida2019,2019NatPh..15.1181K,doi:10.1126/science.aax0274,Liu2020,Kong2021}. 
While MZMs are topologically protected to stay at zero energy (Fermi level) irrespective of local perturbations, their wave functions can still be strongly influenced by disorder. In contrast, CdGM states, lacking topological protection, are expected to respond to disorder in both their energy and wave function.  Despite substantial theoretical progress in understanding average spectral properties~\cite{PhysRevLett.130.106001,RevModPhys.87.1037,PhysRevLett.109.227005,PhysRevB.110.075433,PhysRevLett.132.036604}, distinguishing the statistical behavior of MZMs from CdGM states under generic disorder remains a challenge.

\begin{figure}
\centering
\includegraphics[width=1.0\columnwidth]{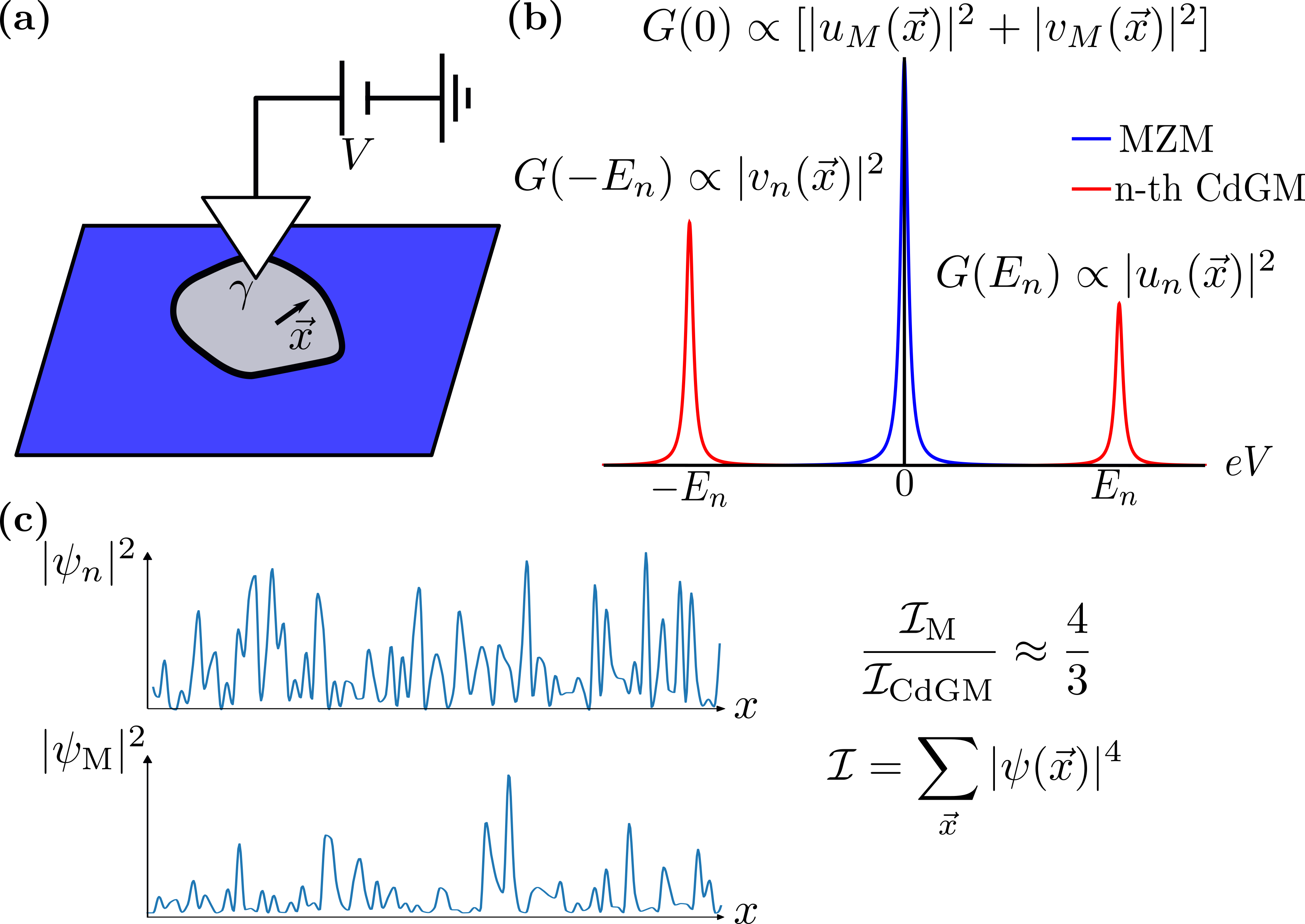}
\caption{ 
STM detection of antidot-bound states. (a) Schematic of the STM setup. The blue and gray regions correspond to the superconducting and normal regions. An STM tip biased at voltage $V$ probes the tunneling conductance inside the antidot, which hosts a MZM $\gamma$. (b) Differential conductance plotted against energy $eV$. The signal maps the thermally broadened LDOS [see Eq.~(\ref{eq:conductance})]. The MZM has a sharp zero-bias peak at $eV=0$. In contrast, the $n$-th CdGM state has two peaks at energies $\pm E_n$. (c) Probability density distribution for the CdGM state and MZM. They exhibit distinct localization statistics; specifically, the ratio of MZM IPR $\mathcal{I}_{\text{M}}$ to the CdGM IPR $\mathcal{I}_{\text{CdGM}}$ is 4/3.
} \label{fig:system}
\end{figure}

To address this challenge in a controlled setting, we focus on a superconducting antidot geometry, realized by removing the superconductor locally in a topological insulator-superconductor heterostructure~\cite{PhysRevLett.100.096407,PhysRevB.89.085409,Deng_2021,Ziesen_2021}, as shown in Fig.~\ref{fig:system}(a). Unlike a conventional Abrikosov vortex, where the normal core size is determined by material parameters, this design offers two advantages: the position of the vortex is spatially fixed, and the geometry allows a pinned flux at a small magnetic field. By stabilizing a vortex hosting a single MZM within this giant vortex setup, we establish an ideal platform to study the impact of disorder on the bound states. 

In this work, we develop a general theory to elucidate the effects of random disorder on a 2D spinless $p_x+ip_y$ superconductor with an antidot. Combining an analytical random matrix theory (RMT) approach with large-scale numerical simulations, we characterize the statistical distributions of both MZMs and CdGM states. We demonstrate that the variance of the probability density exhibits a fundamental difference between the topological MZM and the trivial CdGM states, as shown in Fig.~\ref{fig:system}. This statistical distinctiveness provides a new, spatially resolved criterion for distinguishing MZMs in disordered environments, offering a new way to interpret STM measurements in the antidot system.

\textbf{\emph{Tunneling conductance in STM.~}}

To investigate the low-energy eigenstates in the antidot system, we analyze the differential tunneling conductance, $G(\vec{x},V)\equiv dI/dV$, where $I$ and $V$ are the current and bias voltage between the sample and the tip. This conductance can be directly measured using scanning tunneling microscopy. The STM tip is modeled as a metallic lead weakly coupled to the superconductor surface at a point $\vec{x}$, see  Fig.~\ref{fig:system}(a).  The differential tunneling conductance in the low-temperature limit is a convolution with the local density of states (LDOS) of the antidot~\cite{Tinkham1975IntroductionTS}:
\begin{align}
G(\vec{x},V) &=\frac{G_0\beta}{4\pi\hbar N_1(\mu_0)}\int_0^\infty dE  \cosh^{-2}[\frac{(E-\mu_{0}-eV)\beta}{2}] \label{eq:conductance}\\ 
\int_{-\infty}^\infty  &dE^\prime N_\text{s}(\vec{x},E^\prime) \text{Im} \,
\frac{1}{\frac{E^\prime-E+\mu_{0}}{\hbar}-i\eta}\nonumber,
\end{align}
where $\beta = (k_B T)^{-1}$ is the inverse temperature, $N_1(\mu_0)$ is the normal-state density of states (DOS) of the sample at the Fermi level, and $G_0 = (2\pi e^2/\hbar)|t_{p}|^{2}N_1(\mu_0)N_2(\mu_0)$ is the zero-temperature normal-state conductance determined by the tunneling amplitude $t_p$ and the DOS of the tip $N_2(\mu_0)$. The infinitesimal level width $\eta>0$ ensures causality. The LDOS $N_\text{s}(\vec{x},E_1)$ is determined by the quasiparticle wave functions of the antidot system:
\begin{equation}
N_\text{s}(\vec{x},E)=\sum_{E_n\ge0}|u_n(\vec{x})|^2 \delta(E-E_n)+|v_n(\vec{x})|^2\delta(E+E_n),
\label{eq:LDOS}
\end{equation}
where $\Psi_n(\vec{x})=[u_n(\vec{x}),v_n(\vec{x})]^T$ represents the Nambu spinor eigenstate with energy $E_n$. At low energies $eV, k_B T$ below the superconducting gap, only bound states contribute to the sum. 

The detection of an MZM relies on the spatial and energetic resolution of this conductance, specifically the observation of a zero-bias peak (ZBP) within the vortex core. However, in realistic disordered environments, disorder can induce trivial low-energy states that mimic these spectral features, making the ZBP alone an insufficient criterion. To resolve this ambiguity, we look beyond the local energy spectrum to the spatial statistics of the wavefunctions $u_n, v_n$. We specifically examine the inverse participation ratio (IPR) as a metric to quantify the localization differences between topological and trivial modes.

\textbf{\emph{Random matrix representation.~}}
To capture the universal statistical properties of the antidot bound states in the presence of disorder, we model the effective Hamiltonian projected onto the subspace of low-energy states ($E\ll \Delta$, where $\Delta$ is the pairing potential) as a random matrix, $H^{R}$~\cite{PhysRevLett.130.106001}.  

We analyze an antidot containing a single MZM and $N$ CdGM states. In the Majorana basis, the $(2N+1)$-dimensional random Hamiltonian matrix $H^R$ gives rise to a second quantized Hamiltonian, 
\begin{equation}
\mathcal{H}=\sum_{i,j=1}^{2N+1}H^R_{i,j}\gamma_{i}\gamma_{j},
\label{eq:Majoran-based H}
\end{equation}
where $H^R$ is antisymmetric because of the anticommutation of the Majorana operators $\gamma_i$. Therefore, this odd-dimensional matrix has a zero determinant. Furthermore, the matrix $H^R$ anticommutes with the particle-hole operator $\mathcal{P}=K$, where $K$ is the complex conjugation operator, so  $\mathcal{P}$ maps the state at energy $E$ to a distinct partner at $-E$. Consequently, this effective low-energy Hamiltonian has generically one zero-energy eigenstate (the MZM), $N$ positive-energy eigenstates, and $N$ negative-energy eigenstates. We denote these finite-energy eigenspinors by $\phi_\text{CdGM}$ and the zero energy one by $\phi_{\text{M}}$.

 The distinction between topological and trivial states arises from the fundamental particle-hole symmetry (PHS) of the superconducting system. Consequently, the components of $\phi_\text{CdGM}$~\footnote{$\phi$ is the eigenstate under the Majorana basis while $\psi$ is the eigenstate under Nambu spinor basis.} are complex, obeying statistics consistent with the Gaussian unitary ensemble (GUE)~\footnote{We assume the distribution of the elements of $H^R$ is invariant under unitary transformations. This unitary invariance implies that the eigenvectors of the Hamiltonian are uniformly distributed on the hypersphere of the Hilbert space.}. In contrast, the MZM $\phi_\text{M}$ has a zero eigenvalue, requiring it to be an eigenstate of the PHS operator ($\mathcal{P}\phi_\text{M}=\phi_\text{M}$). This constraint forces the MZM wavefunction to be real in the Majorana basis. Therefore, the MZM vector is restricted to a real submanifold of the Hilbert space, obeying statistics consistent with the Gaussian orthogonal ensemble (GOE). 
This key difference enables one to distinguish these two states in their IPRs, see Eq.~(\ref{eq:LDOS ipr}) below. 

The distinction between real and complex eigenspinors leads to different probability density functions (PDFs) for the squared moduli of any component (first one here for concreteness) $w=|\phi(1)|^2$~\cite{livan2018introduction},
\begin{align}
f_\text{M}(w)&=\frac{1}{\sqrt{\pi}}\frac{\Gamma(D/2)}{\Gamma((D-1)/2)}\frac{(1-w)^{(D-3)/2}}{\sqrt{w}},\label{eq:pdf of MZM moduli}\\
f_\text{CdGM}(w)&=(D-1)(1-w)^{D-2}.
\label{eq:pdf of CdGM moduli}
\end{align}
Here $D=2N+1$ is the dimension of the random matrix $H^R$. We define the disorder average as $\langle \dots \rangle$, which is also the ensemble average over different matrices in Eq.~(\ref{eq:Majoran-based H}). By computing the statistical moments over the ensemble, we find that while the mean intensity $\langle w\rangle$ is identical for both species due to normalization, their fluctuations differ significantly. Specifically, the variance  $\langle (w - \langle w\rangle)^2\rangle$ of the squared wavefunction components exhibits a characteristic difference:
\begin{align}
\text{Var}(w_{\text{M}})&=\frac{2(D-1)}{D^{2}(D+2)}\approx\frac{2}{D^{2}}\quad (\text{at large }D),\label{eq:Variance MZM}\\
\text{Var}(w_\text{CdGM})&=\frac{(D-1)}{D^{2}(D+1)}\approx \frac{1}{D^{2}}\quad (\text{at large }D). \label{eq:Variance CdGM}
\end{align}
Thus the variance for the real-valued MZM is exactly twice that of the complex CdGM states in the large-$D$ limit, providing a statistical observable to distinguish MZM.

\begin{figure}
\centering
\includegraphics[width=1.0\columnwidth]{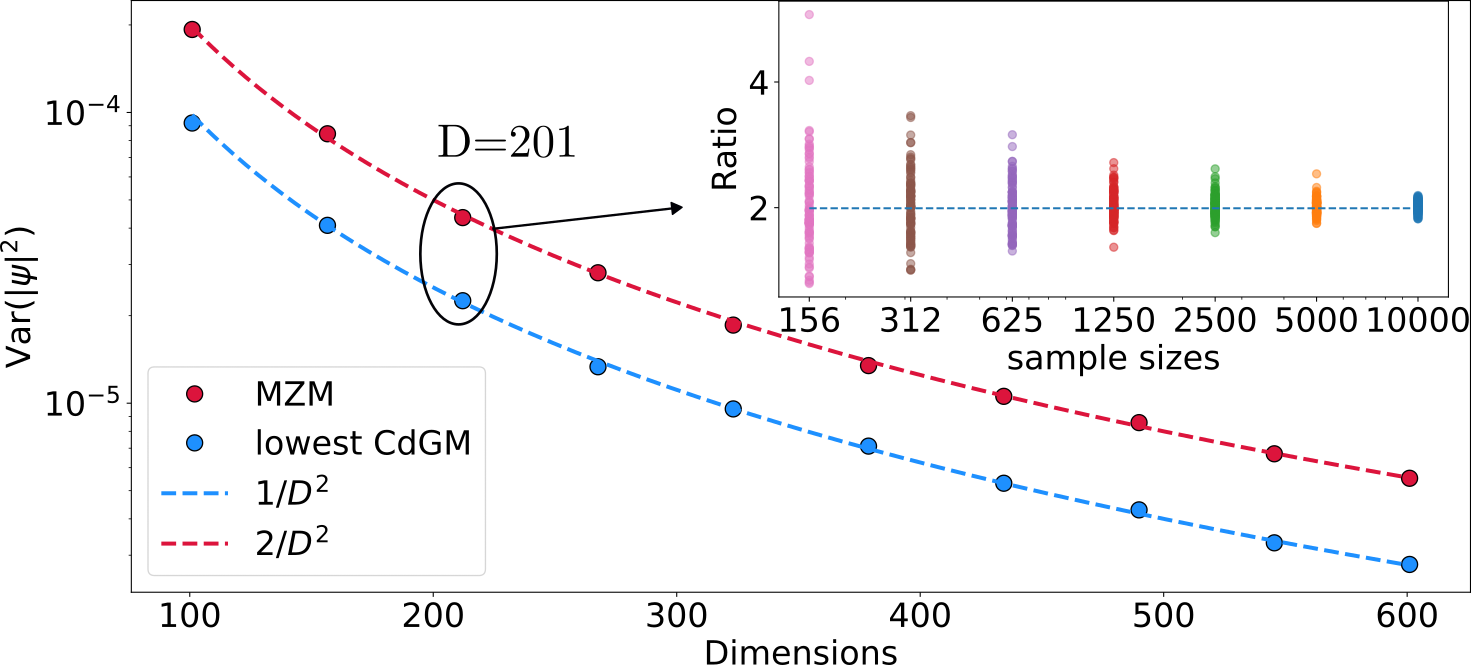}

\caption{ Numerical verification of RMT variance scaling. Main panel: Variance of the squared wavefunction component vs the Hamiltonian dimension $D$. Dots represent simulation results averaged over 20,000 realizations for the MZM and the lowest CdGM, dashed lines indicate the analytical predictions from Eqs.~(\ref{eq:Variance MZM}) and~(\ref{eq:Variance CdGM}). Inset: Convergence of the variance ratio $\text{Var}(\omega_\text{M})/\text{Var}(\omega_\text{CdGM})$ for a fixed dimension $D=201$. The ratio is plotted as a function of subsample size, demonstrating that statistical fluctuations diminish as the sample size increases.
} \label{fig:RMT simulation}
\end{figure}

To verify these analytical predictions, we performed numerical simulations using random matrix ensembles. We varied the dimension of the Hamiltonian $H^R$ from $D=101$ to $D=601$, generating $20,000$ independent realizations for each dimension. For every realization, we extracted the squared modulus of the first spinor component $|\phi(1)|^2$, for both the MZM and the lowest-energy CdGM state. The computed variances scale with dimension $D$ in excellent agreement with our analytical results, as shown in Fig.~\ref{fig:RMT simulation}. The ratio of the variances converges toward the theoretical limit of 2 up to finite-ensemble statistical fluctuations. 
To demonstrate this convergence, the inset of Fig.~\ref{fig:RMT simulation} displays the variance ratio calculated over variable-sized subsamples of the $D=201$ dimension. As the sample size increases, the fluctuations dampen significantly, confirming that the ratio asymptotically approaches the predicted value 2.

\textbf{\emph{Relation of RMT predictions to tunneling conductance.~}}
To relate the RMT predictions to the LDOS, Eq.~(\ref{eq:LDOS}),  measured in tunneling experiments, we transform the system from the Majorana basis to the complex fermion (Nambu) basis. 
The complex fermion operators $c_j^\dagger$ are defined in terms of the Majorana operators as $c_j^\dagger=\frac{1}{2}(\gamma_{2j-1}-i\gamma_{2j})$. 
This relation defines a unitary transformation, $\psi^\text{Nambu}=U\phi^\text{Maj}$, mapping the eigenstates from the Majorana basis to the Nambu spinor basis~\cite{PhysRevB.63.224204,SuppMat}. In the system we consider, there are $2N+1$ coupled Majoranas in the antidot and 1 uncoupled Majorana outside, so the corresponding Nambu spinor is a vector of length $2(N+1)$, $\psi^\text{Nambu}=[u(1),...,u(N+1),v(1),...,v(N+1)]^T$. The transformation matrix $U$ mixes the different components of the Majorana-basis eigenvectors $\phi^\text{Maj}$ into the complex electron ($u$) and hole ($v$) components of the Nambu spinor. This mixing has critical implications for the statistical properties of the wavefunctions in the Nambu basis.

RMT provides the PDFs for the wavefunction components in the Majorana basis [Eqs.~(\ref{eq:pdf of MZM moduli}) and.~(\ref{eq:pdf of CdGM moduli})]. By using the RMT distributions results in Eqs.~(\ref{eq:Variance MZM}) and~(\ref{eq:Variance CdGM}) with the transformation matrix $U$, we derive the statistics of the Nambu components $u$ and $v$ for MZM and CdGMs in the large-$D$ limit~\cite{SuppMat}. We find that the fourth moment for the electron or hole component alone is statistically indistinguishable between the topological and trivial modes:
\begin{align}
\langle |u_{\text{M}}|^4\rangle &\approx \frac{2}{D^2}\approx\langle |u_{\text{CdGM}}|^4 \rangle\label{eq:electron ipr},\\
\langle |v_{\text{M}}|^4\rangle &\approx  \frac{2}{D^2}\approx\langle|v_{\text{CdGM}}|^4 \rangle. \label{eq:hole ipr}
\end{align}
However, the distinction is recovered when considering the total probability density, $\rho(\vec{x})=|u(\vec{x})|^2+|v(\vec{x})|^2$. For MZM, the PHS imposes a strict correlation between the electron and hole components ($|u(\vec{x})|=|v(\vec{x})|$), while for CdGM states, these components are uncorrelated. This correlation difference leads to distinct PDFs for the total density, 
\begin{align}
F_\text{M}(\rho)&=\frac{D-2}{2}(1-\rho)^{\frac{D-4}{2}}\label{eq:pdf of MZM LDOS}\\
F_\text{CdGM}(\rho)&=(D-1)(D-2)\rho(1-\rho)^{D-3}.
\label{eq:pdf of CdGM LDOS}
\end{align}
The variance difference we found in the Majorana basis, Eqs.~(\ref{eq:Variance MZM})-(\ref{eq:Variance CdGM}), translates here into a difference in the second moments of $\rho$, but with a modified coefficient~\cite{SuppMat}:
\begin{align}
\langle \rho_{\text{M}}^2\rangle &\approx \frac{8}{D^2},\qquad \langle\rho_{\text{CdGM}}^2\rangle  \approx  \frac{6}{D^2}.\label{eq:LDOS ipr}
\end{align}
Therefore, we demonstrate that the second moments of the total probability density exhibit a fundamental divergence between the two species:
\begin{equation}
\frac{\langle \rho_{\text{M}}^2\rangle}{\langle\rho_{\text{CdGM}}^2\rangle} \approx \frac{4}{3}.\label{eq:ipr ratio}
\end{equation}
This result establishes that the total probability density provides a disorder-invariant criterion for distinguishing MZMs. To connect this to a spatial observable, we introduce the inverse participation ratio, defined as the spatial summation of the fourth power of the wavefunction, $\mathcal{I}=\sum_{\vec{x}}|\psi(\vec{x})|^4$. Under the assumption of ergodicity, the spatial sum is equivalent to the ensemble expectation value derived from the PDFs~\cite{Frigg_Werndl_2024}. Thus, we have the correspondence:
\begin{equation}
\mathcal{I}=\sum_{\vec{x}}|\psi(\vec{x})|^4\approx N_{\text{site}}\langle |\psi(\vec{x})|^4\rangle =N_{\text{site}}\langle \rho^2\rangle,\label{eq:IPR}
\end{equation}
here $N_{\text{site}}$ is the number of sites in the sum over $\vec{x}$ and the RHS is independent of $\vec{x}$ after the average of the disorder.

Eq.~(\ref{eq:ipr ratio}) is experimentally accessible via the symmetrized conductance, defined as $G_{\text{sym}}(\vec{x},V)=G(\vec{x},V)+G(\vec{x},-V)$ from Eq.~(\ref{eq:conductance}),
\begin{align}
\lim_{\eta \to 0}&\sum_{\vec{x}}G_{\text{sym}}^2(\vec{x},V) \approx[\frac{G_0\beta}{4 \bar{N_1}(\mu_0)}]^2 \sum_{E_n\ge0}\langle \rho_n^2 \rangle \notag\\
&\{\cosh^{-2}[\frac{(E_n-eV)\beta}{2}]+\cosh^{-2}[\frac{(E_n+eV)\beta}{2}]\}^2,  \label{eq:conductance sum}
\end{align}
where $\bar{N_1}(\mu_0)$ is the disorder average of $N_1(\mu_0)$ and we have used the equivalence of spatial and disorder averaging given in  Eq.~(\ref{eq:IPR}). When the voltage is tuned so that tunneling occurs only into either a single CdGM mode $E_n$ or the MZM, the ratio of the symmetrized conductance evaluates to $\lim_{\eta \to 0}\frac{\sum_{\vec{x}}G_{\text{sym}}^2(\vec{x},0)}{\sum_{\vec{x}}G_{\text{sym}}^2(\vec{x},E_n/e)}\approx \frac{\langle \rho_{\text{M}}^2 \rangle}{\langle \rho_n^2 \rangle}$. Consequently, the second moments of $\rho$ can be directly extracted by measuring the spatial second moments of this symmetrized conductance.

\begin{figure}
\centering
\includegraphics[width=1.0\columnwidth]{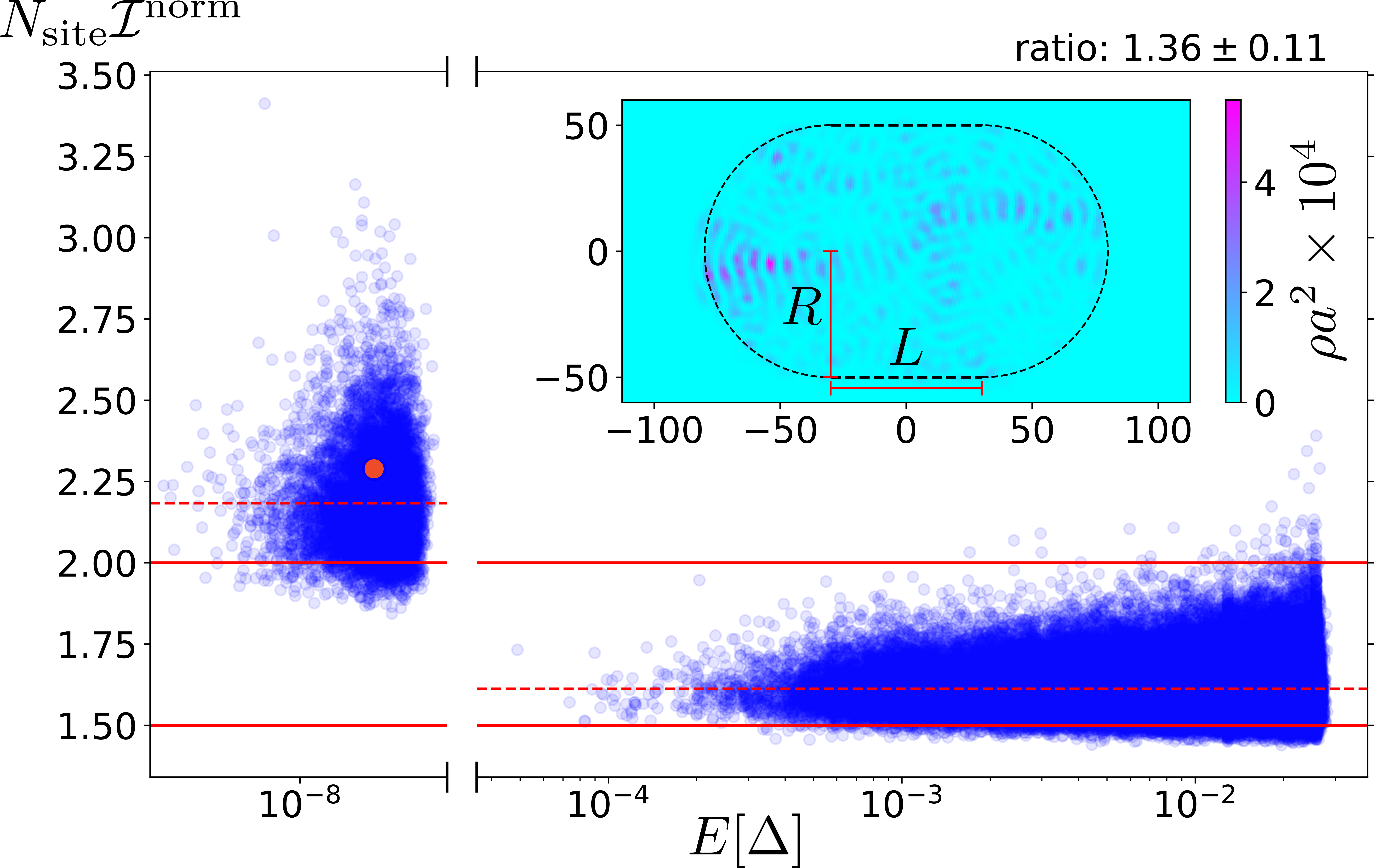}

\caption{Numerical results for the scaled IPR of the density. Main panel: The scaled IPR, $N_{\text{site}}\mathcal{I}^{\text{norm}}$ in Eq.~(\ref{eq:normalized IPR}), plotted versus energy $E$ (in units of $\Delta$). The solid lines indicate the theoretical expected values for MZMs and CdGM states, as given by Eq.~(\ref{eq:LDOS ipr}). The dashed lines represent the mean values averaged over all $10^4$ realizations. The resulting mean ratio of the scaled IPRs is $1.36\pm 0.11$, close to the value predicted by Eq.~(\ref{eq:ipr ratio}). Inset: Density distribution of the single MZM indicated by the red point. The black dashed line outlines the boundary of the stadium antidot. Spatial coordinates are plotted in units of the lattice constant $a$.
} \label{fig:stadium simulation}
\end{figure}

\textbf{\emph{2D spinless p-wave SC simulation~}}
To verify our RMT predictions regarding the ratio Eq.~(\ref{eq:ipr ratio}), we simulate a 2D spinless $p$-wave superconductor on a square lattice. The Hamiltonian is given by $H=\frac{1}{2}\int d\vec{x}\Psi^\dagger H_{\text{BdG}}\Psi$ with~\cite{PhysRevB.105.214521} 
\begin{equation}
H_{\text{BdG}}=\begin{pmatrix}-\frac{1}{2m}\nabla^2-\mu&\Delta \\ \Delta^\dagger&\frac{1}{2m}\nabla^2+\mu&\end{pmatrix}. \label{eq:H_BdG}
\end{equation}
We solve the corresponding tight-binding model using the Kwant package~\cite{groth2014kwant}, with lattice constant $a=25$~nm, chemical potential $\mu=100$~{\textmu}eV, and the hopping amplitude $t=\frac{\hbar^2}{2m a}\approx 400$~{\textmu}eV. The system is a  $240a \cross 240a $ square, the bulk $p_x+ip_y$ pairing potential $\Delta$ vanishes inside the antidot and takes the form $2a\Delta_0e^{i\theta}(\partial_x +i\partial_y)$ outside, where $\Delta_0=100$~{\textmu}eV. To induce quantum chaos, we introduce a stadium-shaped antidot with length  $L=60a$ and radius $R=50a$, displayed in the inset of Fig.~\ref{fig:stadium simulation}.

\begin{figure}
\centering
\includegraphics[width=1.0\columnwidth]{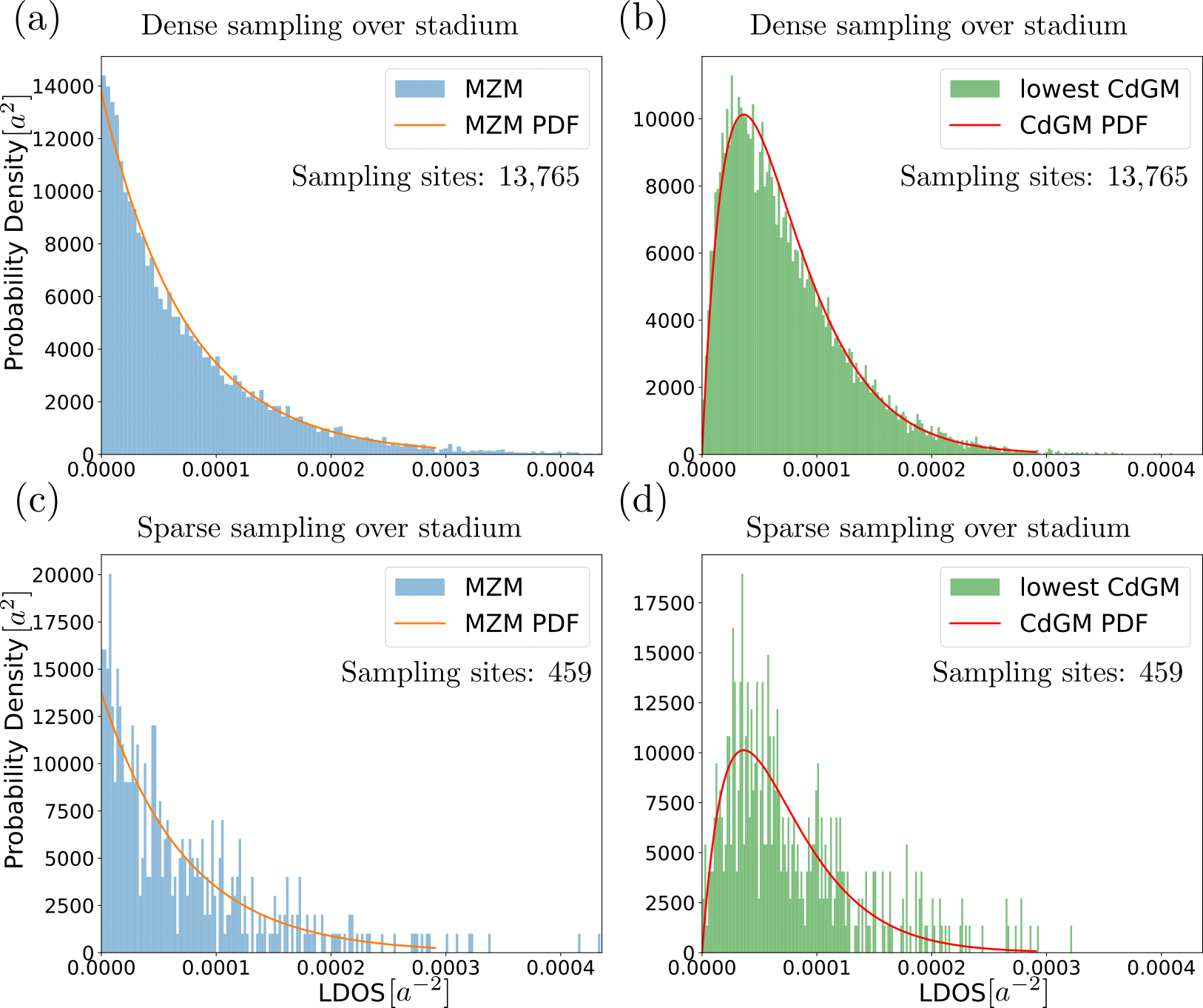}

\caption{Histograms representing the statistical distributions of the total probability density $\rho(\vec{x})$, in a stadium antidot, calculated numerically for a single random disorder realization. (a-b) The MZM (a) and lowest CdGM (b) sampled over the whole stadium $\mathcal{S}$   (13,765 sampling sites). The CdGM-to-MZM total probability density variance ratio is 1.83, approaching the theoretically expected value of 2. (c-d) The MZM (c) and the lowest CdGM (d) within a uniform sub-sampled region of the stadium comprising 459 sampling sites. The CdGM-to-MZM total probability density variance ratio is 1.29. The solid orange and red curves represent the theoretical PDFs for the MZM and CdGM states, as given in Eqs.~(\ref{eq:pdf of MZM LDOS}) and (\ref{eq:pdf of CdGM LDOS}), respectively.
} \label{fig:stadium pdf}
\end{figure}

However, in a clean ballistic stadium, spatial symmetries prevent the system from exhibiting expected random matrix statistics~\cite{Bies_2001,Batistic_2020}. To break these symmetries and ensure generic chaotic behavior, we introduce onsite disorder modeled via a position-dependent random potential drawn from a uniform distribution over $[-\Delta_0,\Delta_0]$. We analyze the 20 lowest positive-energy states across $10^4$ independent disorder realizations. We define the normalized IPR within the stadium region $\mathcal{S}$ as:
\begin{equation}
\mathcal{I}^{\text{norm}}(\rho)=\frac{\sum_{\vec{x}\in \mathcal{S}}\rho(\vec{x})^2}{[\sum_{\vec{x}\in \mathcal{S}}\rho(\vec{x})]^2}.\label{eq:normalized IPR}
\end{equation}
Since the Hamiltonian, Eq.~(\ref{eq:H_BdG}), is a $2\cross 2$ matrix, the effective low-energy subspace dimension $D$ is twice the number of sites in the antidot $N_{\text{site}}$. Consequently, Eq.~(\ref{eq:LDOS ipr}) predicts the scaled IPR values $ N_{\text{site}}\mathcal{I}^{\text{norm}} (\rho_\text{M})\approx 2$ and $N_{\text{site}}\mathcal{I}^{\text{norm}} (\rho_\text{CdGM})\approx 3/2$. Our numerical results in Fig.~\ref{fig:stadium simulation} yield an average ratio between these normalized IPRs of 1.36, showing agreement with the theoretical prediction of 4/3 in Eq.~(\ref{eq:ipr ratio}). Furthermore, we verify in Fig.~\ref{fig:stadium pdf} that the total probability density PDFs for a single disorder realization agree with the theoretical distributions provided in Eqs.~(\ref{eq:pdf of MZM LDOS}) and (\ref{eq:pdf of CdGM LDOS}).

\textbf{\emph{Discussion~}}
We studied an antidot vortex hosting a Majorana zero mode and Caroli-de Gennes-Matricon states. 
In the right basis, the wave function of the former is real while the latter has a complex wave function. 
This property remains the only difference between the wave functions of the MZM and CdGM states in a disordered antidot described by random matrix theory. 
We showed that the difference is observable in the statistical properties (such as the IPR) of the total probability density that can be probed in STM measurements. We predict the ratio of  IPRs to be 4/3, see Eq.~(\ref{eq:ipr ratio}). 

The required statistical ensemble can be obtained from a single device. In practice, STM measurements taken at different positions inside the antidot provide total probability density distributions such as shown in Fig.~\ref{fig:stadium pdf}. Under the optimal binary hypothesis test, the Chernoff information~\cite{Chernoff1952AMO,cover2006elements} sets an information-theoretic lower bound of 83 measurements to distinguish the two total probability density distributions (MZM vs CdGM) at a $< 5\%$  error rate. (Fig.~\ref{fig:stadium pdf} shows larger ensembles to resolve the full PDFs distributions more clearly.) Alternatively, different disorder realizations can be generated by varying a gate voltage applied to the antidot~\cite{PhysRevLett.130.106001}. 

Each spatial sampling region must be larger than the relevant disorder correlation length, typically of order the mean free path, so that the total probability density measurements are statistically independent. This condition sets a minimum size for the antidot. 
Larger antidots host more states, driving the system closer to the universal RMT limit and ensuring fully developed chaotic dynamics. However, increasing the antidot size reduces the level spacing, lowering the energy of the CdGM states. Consequently, resolving the zero-energy MZM from the lowest CdGM state requires spectral resolution finer than the lowest CdGM energy.

Poor spectral resolution causes the STM tip to admix MZM and CdGM contributions; the measured IPR is then a weighted sum of the two contributions. The resulting IPR interpolates between the MZM and CdGM limiting values and should not exceed the universal upper bound of $4/3$. Equal admixture yields a weighted average $7/6$.  These limits are experimentally testable benchmarks.

\begin{acknowledgments} 
\textbf{\emph{Acknowledgments.~}} 
We thank P.~W.~Brouwer, Y.~P.~Chen, M.~Fuhrer, Y.~Gefen, and I.~Giorgadze for valuable discussions.  
This material is based upon work supported by the Office of the Under Secretary of Defense for Research and Engineering under award number FA9550-22-1-0354. 
This work has been supported by the U.S. Department of Energy, Office of Science, National Quantum Information Sciences Research Centers, Quantum Science Center.
\end{acknowledgments}

\bibliography{refs}
\clearpage
\newpage

\setcounter{equation}{0}
\setcounter{figure}{0}
\setcounter{section}{0}
\setcounter{table}{0}
\setcounter{page}{1}
\makeatletter
\renewcommand{\theequation}{S\arabic{equation}}
\renewcommand{\thepage}{S-\arabic{page}}
\renewcommand{\thesection}{S\arabic{section}}
\renewcommand{\thefigure}{S\arabic{figure}}
\renewcommand{\thetable}{S-\Roman{table}}

\begin{widetext}
\begin{center}
Supplementary materials on \\
\textbf{``Statistical Signatures of Majorana Zero Modes in Disordered Topological Superconductor Antidot Vortices''}\\
Zhibo Ren$^{1}$, Jukka I. V{\"a}yrynen$^{1}$, \\ 
$^{1}$ \textit{Department of Physics and Astronomy, Purdue University, West Lafayette, Indiana 47907, USA}\\
\end{center}

These supplementary materials contain details about IPR in the Nambu basis.

\section{Statistics in the Nambu basis}
With the definition of the fermion creation operator $c_j^\dagger=\frac{1}{2}(\gamma_{2j-1}-i\gamma_{2j})$, we can construct the unitary transformation matrix $U$ between the Nambu spinor $C=(c_{1},\ldots,c_{N+1},c_{1}^{\dagger},\ldots,c_{N+1}^{\dagger})^{T}$ and the Majorana spinor $\gamma=(\gamma_{1},\ldots\gamma_{2N+2})^{T}$, such that $C=\frac{\sqrt{2}}{2}U \gamma$. Here, $\gamma_{2N+2}$ denotes the uncoupled Majorana operator outside the antidot. The matrix $U$ takes the form
\begin{equation}
U=\frac{\sqrt{2}}{2}\left(\begin{array}{cccc}
1 & i& 0 & 0\cdots\\
0 & 0 & 1 & i\cdots\\
\vdots & \vdots & \vdots & \vdots\\
1 & -i& 0 & 0\cdots\\
\vdots & \vdots & \vdots & \vdots\\
\end{array}\right).
\label{eq:unitary matrix}
\end{equation}
We also have $\psi^\text{Nambu}=U\phi^\text{Maj}$. Denoting the Nambu eigenstate as $\psi_n^{\text{Nambu}}=(u_n,v_n)^T$, where $u_n$ and $v_n$ are the electron and hole components, respectively, the first components of these wave functions can be expressed in terms of the Majorana-basis eigenstate as
\begin{equation}
u_n(1)=\frac{\sqrt{2}}{2}\big[\phi_n(1)+i\phi_n(2)\big],\qquad
v_n(1)=\frac{\sqrt{2}}{2}\big[\phi_n(1)-i\phi_n(2)\big].\label{eq:u and v}
\end{equation}
The first component of the total probability density $\rho_n$ is then \begin{equation}
\rho_n(1)=|u_n(1)|^2+|v_n(1)|^2
=|\phi_n(1)|^2+|\phi_n(2)|^2.
\label{eq:LDOS form}
\end{equation}
Defining $w^{(1)}=|\phi(1)|^2$ and $w^{(2)}=|\phi(2)|^2$, we obtain the joint probability distributions for MZM and CdGM states from RMT:
\begin{align}
f_\text{M}(w^{(1)},w^{(2)})&=\frac{1}{\pi}\frac{\Gamma(D/2)}{\Gamma(D/2-1)}\frac{[1-(w^{(1)}+w^{(2)})]^{(D-4)/2}}{\sqrt{w^{(1)}w^{(2)}}},\\
f_\text{CdGM}(w^{(1)},w^{(2)})&=(D-1)(D-2)[1-(w^{(1)}+w^{(2)})]^{D-3}.
\label{eq:joint pdf of eigenstates}
\end{align}
In the large-$D$ limit, $w^{(1)}$ and $w^{(2)}$ can be treated as effectively independent random variables, with distributions given in Eqs.~(\ref{eq:pdf of MZM moduli}) and~(\ref{eq:pdf of CdGM moduli}).

The modulus of the first components of the electronic wavefunction is
\begin{equation}
|u_n(1)|^2=\frac{1}{2}\Big[w^{(1)}+w^{(2)}+2\,\Im\!\big(\phi_n(1)\phi_n^*(2)\big)\Big].
\label{eq:elctron part moduli}
\end{equation}
Since the phases of $\phi_n(1)$ and $\phi_n(2)$ are completely random, the expectation value of $\Im\!\big(\phi_n(1)\phi_n^*(2)\big)$ vanishes, whereas the expectation value of its square is nonzero, $\big\langle \Im^2\!\big(\phi_n(1)\phi_n^*(2)\big)\big\rangle
=\frac{1}{2}\langle w^{(1)}w^{(2)}\rangle$. Using the independence condition together with the distributions in Eqs.~(\ref{eq:pdf of MZM moduli}) and~(\ref{eq:pdf of CdGM moduli}), we obtain
\begin{equation}
\langle |u_{\text{M}(1)}|^4\rangle \approx \frac{2}{D^2}\approx\langle |u_{\text{CdGM}}(1)|^4 \rangle\label{eq:electron ipr2},\\
\end{equation}
which is the result quoted in Eq.~(\ref{eq:electron ipr}) of the main text.
An analogous result holds for the hole component $v_n$, so the statistics of the electron and hole amplitudes alone do not distinguish MZMs from CdGM states.

We now turn to the total probability density. Since $\rho_n(1)=w^{(1)}+w^{(2)}$, the PDF of $\rho_n(1)$ in Eqs.~(\ref{eq:pdf of MZM LDOS}) and~(\ref{eq:pdf of CdGM LDOS}) can be obtained via the convolution integral $F[\rho_n(1)]=\int_{0}^{\rho_n(1)}d\omega f(\omega ,\rho_n(1)-\omega)$, which yields
\begin{align}
F_\text{M}[\rho_n(1)]&=\frac{D-2}{2}[1-\rho_n(1)]^{\frac{D-4}{2}},\label{eq:pdf of LDOS MZM}\\
F_\text{CdGM}[\rho_n(1)]&=(D-1)(D-2)\rho_n(1)[1-\rho_n(1)]^{D-3}.
\label{eq:pdf of LDOS CdGM}
\end{align}
The corresponding first and second moments are
\begin{align}
\langle \rho_{M}(1)\rangle=\frac{2}{D}&,\qquad\langle \rho_{M}^{2}(1)\rangle=\frac{8}{D(D+2)}\approx\frac{8}{D^{2}} ,\label{eq:LDOS of MZM}\\
\langle \rho_{CdGM}(1)\rangle=\frac{2}{D}&,\qquad\langle \rho_{CdGM}^{2}(1)\rangle=\frac{6}{D(D+1)}\approx\frac{6}{D^{2}}.
\label{eq:LDOS of CdGM}
\end{align}
These results lead directly to Eqs.~(\ref{eq:LDOS ipr}) and~(\ref{eq:ipr ratio}) of the main text.

\end{widetext}
\end{document}